\documentclass[aps,prl,twocolumn,floatfix,showpacs]{revtex4}
\usepackage{graphics}

\begin{document}

\title{Toward one-band superconductivity in MgB$_2$}
\author{Steven C. Erwin and I.I. Mazin}
\affiliation{Center for Computational Materials Science, Naval Research Laboratory,
Washington, D.C. 20375}
\date{\today}

\begin{abstract}
The two-gap model for superconductivity in MgB$_{2}$ predicts that
interband impurity scattering should be pair breaking, reducing the
critical temperature.  This is perhaps the only prediction of the
model that has not been confirmed experimentally.  It was previously
shown theoretically that common substitutional impurities lead to
negligible interband scattering---if the lattice is assumed not to
distort. Here we report theoretical results showing that certain
impurities can indeed produce lattice distortions sufficiently large
to create measurable interband scattering. On this basis, we predict
that isoelectronic codoping with Al and Na will provide a decisive
test of the two-gap model.
\end{abstract}

\pacs{74.62.Dh,74.70.Ad,61.72.Bb,61.72.Ji}
\maketitle

It is now widely accepted that MgB$_{2}$ is a two-gap
superconductor: its Fermi surface consists of two distinct sheets
characterized by strong and weak electron-phonon coupling,
respectively (see Ref.~\onlinecite{mazin03} for a review). This view
is supported by numerous experiments probing either the larger or
smaller gap, or both simultaneously. Experimental observation of the
merging of the two gaps would constitute even stronger evidence. Such
a merging is expected, for example, from interband scattering by
impurities. Within the theory of multiband superconductivity,
interband scattering mixes the \textquotedblleft
weak\textquotedblright\ and \textquotedblleft
strong\textquotedblright\ Cooper pairs, averaging the order parameter
and reducing $T_{c}$ \cite{golubov97}. At small defect concentrations
the suppression of $T_{c}$ should be linear, with the larger gap
decreasing and the smaller gap increasing.  The effect should be
pronounced in samples with high defect concentrations, but despite the
relatively low quality of many samples such an effect has not been
observed. Indeed, some samples with high resistivity have nearly the same
critical temperatures as clean single crystals.  On the other hand, samples
doped with carbon show both gaps decreasing but not merging, despite
a considerable reduction in $T_{c}$. Other types of intentional
defects---whether from doping or irradiation---have also
failed to produce a merged gap.  Even efforts to introduce defects
into MgB$_{2}$ for the explicit purpose of inducing interband
scattering and merging the two gaps have failed to observe this
effect \cite{wang02,schmidt03,samuely03,balaselvi03}. 

This apparent lack of evidence for a central prediction of the
two-gap model is disturbing.  In Ref.~\cite{mazin02} it was shown that
interband scattering from substitutional impurities is inherently
weak, {\em if the lattice is assumed not to distort.} This partially
explains the null results of current experimental efforts to induce pair
breaking, but it does not address the possibility of pair breaking from
impurities specifically chosen to maximize the interband scattering
due to large lattice distortions.  In this Letter we show that
this strategy is likely to succeed and, by identifying a simple
relationship between impurity atoms and the resulting lattice
distortions in MgB$_2$, suggest an impurity-doping protocol that will
produce measurable pair breaking---and thereby provide the
final ``smoking-gun'' evidence for the two-band model.

Hampering such investigations is the currently limited insight into
which defects are most effective in creating interband
scattering. Since the states at the Fermi level of MgB$_{2}$ are
formed by the boron orbitals, one might expect impurities (such as
carbon) in the B planes to produce large interband scattering. This is
not borne out experimentally: substitutional C impurities have
only a weak effect on the interband scattering \cite{samuely03}. This
finding had been anticipated theoretically as a consequence of the
special symmetry properties of the electronic states within the
$\sigma $ band near the $\Gamma$ point \cite{mazin02}. The crucial
point is that although impurities in the B plane do have a strong
effect on the electronic structure, they do not change the local point
symmetry and therefore do not lead to significant $\sigma -\pi $
scattering.

The situation can be quite different for substitutions in the Mg
plane, which may create out-of-plane distortions of B atoms in
neighboring planes. Such relaxations do change the local point
symmetry of nearby B atoms, mixing the in-plane $p_{x,y}$ and
out-of-plane $p_{z}$ orbitals, and for sufficiently large disturbances
can lead to significant $\sigma -\pi $ scattering. Here we demonstrate
by first-principles calculations that this is indeed the case for
certain substitutional impurities but---surprisingly---not for Mg
vacancies. We predict that the interband scattering effects will be
most pronounced for isoelectronic co-doping with Na and Al, and that
the effect on the superconducting properties should be detectable for
impurity concentrations above 2\%.

We used density-functional theory (DFT) to study the lattice
distortion created by Mg-plane substitutional impurities from Groups
I, II, and III, by a Mg vacancy, and by B-plane C substitution. To
model the distortion induced by single defects, we used 2$\times
$2$\times $2 supercells of bulk MgB$_{2}$ (and 3$\times $3$\times $3
supercells for convergence checks). Total energies and forces were
calculated using projector-augmented-wave potentials and the
generalized-gradient approximation \cite {kresse93a,kresse96a}. All
atomic positions were relaxed within the constraint of fixed
(theoretical) bulk lattice parameters. The resulting displacements of
the nearest-neighbor B atoms are given in Table I. Given the strong
intraplanar covalent bonding, it is not surprising that these
displacements are dominated by the out-of-plane component $\delta z$,
which ranges from $-$0.04 \AA\ (for Be) to $+$0.09
\AA\ (for K); for C it is zero by symmetry. 

\begin{table}
\caption{First-principles displacements of 
nearest-neighbor B atoms, in angstrom, induced by various
substitutional impurities. Positive values indicate displacements away
from the impurity.}
\begin{ruledtabular}
\begin{tabular}{lccc}
Impurity & Site & In-plane, $\delta r$ & Out-of-plane, $\delta z$ \\
\hline
   Be  &  Mg  &     $-$0.014  &       $-$0.039 \\
   Al  &  Mg  &     $-$0.012  &       $-$0.033 \\
   Sc  &  Mg  &     $+$0.012  &       $-$0.008 \\
    C  &  B   &     $+$0.044  &          0     \\
   Li  &  Mg  &     $-$0.006  &       $+$0.003 \\
Vacancy&  Mg  &     $-$0.005  &       $+$0.008 \\
   Ca  &  Mg  &     $+$0.015  &       $+$0.028 \\
   Na  &  Mg  &     $+$0.008  &       $+$0.040 \\
    K  &  Mg  &     $+$0.024  &       $+$0.092 \\
\end{tabular}
\end{ruledtabular}
\end{table}

The out-of-plane displacements for seven different impurities on the
Mg site are plotted in Fig.~1. In a previous related study, changes in
interlayer spacing induced by the complete substitution of Al for Mg
in a single plane were ascribed to electrostatic effects
\cite{barabash02}. We find no such correlation between $\delta z$ and
the formal valence of the impurity---for example, Be and Ca give
displacements of opposite sign despite having identical valence---and
thus infer that electrostatic effects are not important. On the other
hand, Fig.~1 shows that there is an excellent correlation between
$\delta z$ and the ionic radius of the impurity atom.  Hence, we
conclude that the out-of-plane displacement of B atoms by impurity
atoms substituting for Mg is mostly a size effect.

In light of this, one might anticipate a Mg vacancy to produce a large
inward displacement. Our DFT results reveal very different behavior:
the vacancy creates a negligibly small displacement, $\delta z<0.01$
\AA . This result is consistent with the experimental fact that pair
breaking is not observed in low quality samples, which presumably
contain many vacancies.  However, it is very different from the trend
shown in Fig.~1. Indeed, if one naively considers the vacancy as an
impurity of zero size, the predicted displacement is almost twice that
found for Be and Al, in sharp distinction to our DFT result.

A simple model explains the surprisingly small displacement
created by the Mg vacancy. We consider a MgB$_2$ crystal containing a
single substitutional defect $D$ (either an impurity atom or a
vacancy) in the Mg plane. Such a defect has 12 nearest-neighbor B
atoms, consisting of two hexagonal rings.  We consider the
out-of-plane displacement of the B atoms in these rings to arise from
two opposing effects.  The first of these represents the change in
covalent bonding {\em between} B planes.  In the absence of any
defects, the MgB$_2$ interlayer spacing $c_{\rm Mg}$ is primarily
determined by assisted hopping between B $p_z$ orbitals through Mg
orbitals; for each B atom there are three such hopping paths through
nearest-neighbor Mg atoms.  With the defect present, one of these
three paths now passes through the defect site. This new hopping path
results in an out-of-plane force on the B atom. We assume the
magnitude of this effect to be one-third of that found for a fully
substituted Mg plane, which we approximate using the energy vs. layer
spacing, $E_D(c)$, for fully substituted $D$B$_2$. Thus, we consider
the change in spacing between the two displaced hexagonal rings,
$2\delta z$, to contribute an energy per B given by
$\frac{1}{3}E_D(c_{\rm Mg}+2\delta z)$. We have calculated the
binding-energy curves $E_D(c)$ within DFT for AlB$_2$, NaB$_2$, and
``vacancy-substituted'' B$_2$; the results are shown in Fig.~2.  All
can be accurately represented by a Morse potential, which is the form
we will use in the discussion below.

\begin{figure}[tbp]
\resizebox{7.5cm}{!}{\includegraphics{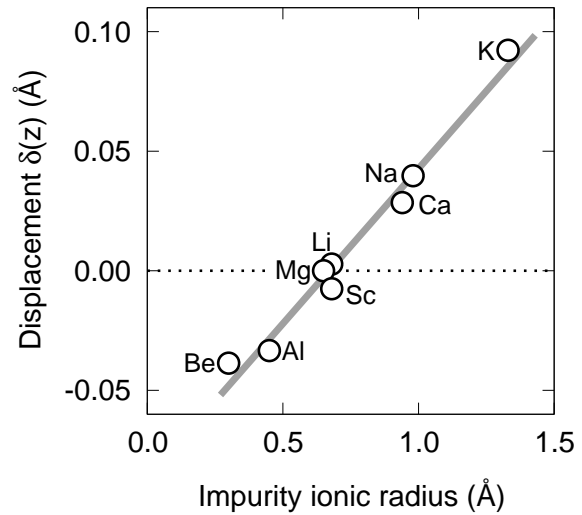}}
\caption{First-principles theoretical displacements of nearest-neighbor B
atoms around a substitutional impurity in the Mg plane. Ionic radii
are from Ref.~\onlinecite{kittel76}.}
\end{figure}

The second effect is the restoring force experienced by the displaced
B atoms, due to the strong covalent bonding {\em within} the B
planes. For the small displacements we are considering it is
reasonable to take this effect to be harmonic in $\delta z$, again
weighted by 1/3. Thus we take the total energy change per B to be
\begin{equation}
E(\delta z)=\frac{1}{3}K(\delta
z)^{2}+\frac{1}{3}kw^{2}[1-e^{-(c_{\mathrm{Mg}}+2\delta
z-c_{D})/w}]^{2},
\end{equation}
where $w$ is the width of the Morse potential, $kw^{2}$ its depth, and
$c_D$ its equilibrium interlayer spacing.  For small
$c_{\mathrm{Mg}}-c_{D}$, it is easy to show that this energy is
minimized for $2\delta z=(c_{D}-c_{\mathrm{Mg}})/(1+K/2k)$. In other
words, for impurities whose size is comparable to Mg, the displacement
is linear in the size mismatch and reduced by the factor $1+K/2k$
(which is typically in the range 5--10); this is consistent with the
DFT results shown in Fig.~1.  Qualitatively, when $c_D$ is close to
$c_{\rm Mg}$, as it is for AlB$_2$ and NaB$_2$, the equilibrium
displacement represents a balance between the harmonic restoring force
$\frac{2}{3}K|\delta z|$ and a nearly harmonic (attractive or
repulsive) force $\frac{2}{3}k(c_{\rm Mg}-c_D+2\delta z)$ from the
interlayer bonding.

For the vacancy there is a very large size mismatch:
$c_{\mathrm{vac}}$ is over 40\% smaller than $c_{\mathrm{Mg}}$. Thus for any
reasonable displacement, the hexagons experience only a weak
attractive force from the tail of the Morse potential.  Hence, the
energy is minimized for a very small displacement, $\delta z\approx
-0.007$ \AA , in agreement with the negligible displacement given by
DFT. Qualitatively, the result of a large size mismatch is to largely
preempt the mechanism of interlayer binding, leading to very small
displacements strongly suppressed by the penalty for perturbing the
planarity of the B layer.

The origin of the large mismatch between $c_{\mathrm{Mg}}$ and
$c_{\mathrm{vac}}$ can best be understood by comparing the band
structures of the fully substituted $D$B$_2$ compounds at their
equilibrium interlayer spacings.  For these pure compounds
(as well as the parent material) interlayer bonding arises primarily
from the interaction between B $p_{z}$ orbitals in different layers.
This interaction depends on the assisted hopping through $s$ and $p_z$
orbitals located on the $D$ site.  However, for the fully ``vacancy
substituted'' compound the occupancy of the $\pi$ bands is so much
reduced that their contribution to bonding becomes quite small. At the
same time, the $\sigma$ bands acquire substantial $z$-dispersion from
$pp\pi$ hopping, which contributes to bonding. Even for high vacancy
concentrations (without complete removal of a Mg plane) it is
impossible to engage the $\sigma$ bands in interlayer bonding by any
reasonable dimpling of the B planes, because the planes remain too far
apart. As a result, hardly any distortion occurs at all: indeed, even
for 50\% vacancies within a single Mg plane, the interlayer spacing
changes by less than 0.05 \AA.

\begin{figure}[tbp]
\resizebox{7cm}{!}{\includegraphics{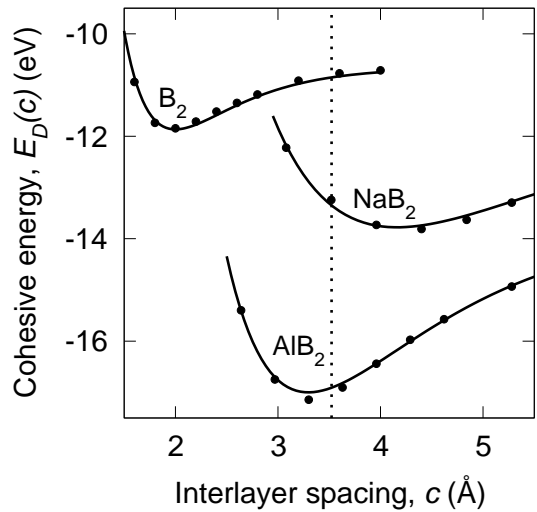}}
\caption{First-principles cohesive energy of completely substituted
MgB$_2$, versus interlayer spacing.
Curves are fits to Morse potentials. The vertical dotted line marks
the equilibrium interlayer spacing for MgB$_2$.}
\end{figure}

We have established that the defects leading to the largest B
displacements are Mg-plane substitutional impurities with a large size
mismatch (but not Mg vacancies).  We now estimate the
magnitude of the interband scattering associated with displacements
from such impurities. We assume that the only relevant scattering is
that due to the out-of-plane distortion, $\delta z$, and use the
analogy between the formulas \cite{allen78} for the impurity-induced
scattering rate,
\begin{equation}
\gamma _{\mathrm{imp}}=\pi n_{\mathrm{imp}}\frac{\sum_{\mathbf{kk}^{\prime
}}\delta (\varepsilon _{\mathbf{k}})\delta (\varepsilon _{\mathbf{k}^{\prime
}})|\delta V_{\mathbf{kk}^{\prime }}|^{2}}{\sum_{\mathbf{k}}\delta
(\varepsilon _{\mathbf{k}})},  \label{imp}
\end{equation}%
and the electron-phonon coupling constant, 
\begin{equation}
\lambda =\frac{\sum_{\nu ,\mathbf{kk}^{\prime }}\delta (\varepsilon _{%
\mathbf{k}})\delta (\varepsilon _{\mathbf{k}^{\prime }})|M_{\nu ,\mathbf{kk}%
^{\prime }}|^{2}/\hbar \omega _{\nu ,\mathbf{k-k}^{\prime }}}{\sum_{\mathbf{k%
}}\delta (\varepsilon _{\mathbf{k}})}.  \label{lam}
\end{equation}%
Here $n_{\mathrm{imp}}$ is the impurity concentration; $\delta
V_{\mathbf{kk}^{\prime }}$ is the matrix element of the impurity
perturbation potential (defined as the difference between the full
crystal potentials with and without an impurity); $\varepsilon
_{\mathbf{k}}$ is the electron energy with respect to the Fermi level;
$\omega _{\nu \mathbf{k}}$ is the phonon frequency;
$M_{\mathbf{k,k}^{\prime }}$ is the electron-ion matrix 
element $\left\langle \mathbf{k|}dV/dQ_{\nu }\mathbf{|k}^{\prime }\right\rangle $,
where $dV/dQ_{\nu }$ is the derivative of the crystal potential with
respect to the phonon normal coordinate $Q=\sqrt{2m\omega /\hbar
}\,x$; and the summations are over all electron states and all phonon
branches \cite{allen78}.

To proceed we make three approximations, all qualitatively reasonable
if not quantitatively reliable. First, we assume that in-plane
phonons, including the well-known $E_{2g}$ modes, contribute little to
\textit{inter}band electron-phonon coupling; this follows from the
same symmetry arguments given in Ref.~\onlinecite{mazin02}. Second, we
assume by the same reasoning that interband impurity scattering comes
only from the out-of-plane relaxation of B atoms. Finally, we
approximate Eq.~\ref{imp} as
\begin{equation}
\gamma _{\mathrm{imp}}\approx 12\pi n_{\mathrm{imp}}\left( \sum_{\mathbf{k}%
}\delta (\varepsilon _{\mathbf{k}})\right) \left\langle \left\vert
dV/du_{z}\right\vert ^{2}\right\rangle (\delta z)^{2},
\end{equation}%
and likewise Eq.~\ref{lam} as 
\begin{equation}
\lambda \approx 2\left( \sum_{\mathbf{k}}\delta (\varepsilon _{\mathbf{k}%
})\right) \left\langle \left\vert dV/du_{z}\right\vert ^{2}\right\rangle
/2m\omega ^{2}.
\end{equation}%
Here we have assumed that the average of the crystal potential with respect
to the vertical displacement of the B atom is the same in both cases. The
numerical factors 12 and 2 are the coordination of Mg and the number of B
atoms in the unit cell, respectively.

First-principles calculations give the interband part of the
electron-phonon coupling as $\sim $0.2 \cite{mazin03}. The phonon
frequencies for the out-of-plane modes are about 400 cm$^{-1}.$ Using
these values, the scattering rate is given by $\gamma
_{\mathrm{imp}}\approx 50\,n_{\mathrm{imp}}(\delta z)^{2}$ eV, where
$\delta z$ is in angstrom. Hence, for 2\% Al doping we find $\gamma
_{\mathrm{imp}}\approx 1.1$ meV.  For 2\% Na doping we estimate a
similar scattering rate, $\gamma _{\mathrm{imp}}\approx 1.3$ meV.
These scattering rates are small, but should still have a measurable
effect on the superconducting gaps and temperature. The effect on the
gaps is difficult to estimate without full Eliashberg calculations.
The reduction of $T_{c}$ can be easily estimated using Eq.~13 from
Ref.~\cite{golubov97}, which gives 2.0--2.5 K (in addition to any
suppression due to the electron doping of the $\sigma$ band).  While
this is a small reduction compared to the changes observed in heavily
electron-doped samples, the underlying mechanism is quite
different. In particular, pair breaking from interband scattering is
unique in that it reduces $T_c$ and the gap ratio while simultaneously
increasing the smaller gap \cite{golubov97}. This distinctive behavior
should facilitate the separation of interband scattering from other
sources of $T_c$ reduction.

Finally, we suggest that an especially attractive test of these
predictions would be simultaneous codoping by equal parts Al and
Na. This would effectively be an isoelectronic substitution, and any
effect on the superconducting properties could then be ascribed to
impurity-induced interband scattering. Moreover, Na and Al induce
distortions of the same magnitude but of the opposite sign, which
should mitigate the effects of a reduction in scattering due to
possible short-range ordering of the impurities.

In conclusion, we have performed first-principles calculations of the
lattice distortion in MgB$_{2}$ induced by several common
substitutional (for Mg) impurities, and a Mg vacancy. We find
out-of-plane displacements as large as 0.04 \AA\ for common impurities
such as Al and Na. The magnitude and sign of the displacement are
mainly determined by the ionic size of the impurity. On the other
hand, for the Mg vacancy we find an essentially negligible
displacement of nearby B atoms. The different behavior of vacancies
and impurities is explained by a simple physical model representing
the competition between interlayer binding and intralayer
planarity. We estimate the interband scattering rate due to the Na and
Al impurities to be of order 1 meV, sufficiently large to give a
detectable change in the superconducting transition
temperature. Finally, we propose that the codoped material
Mg$_{1-x}$Na$_x$Al$_x$B$_2$, which is isoelectronic with MgB$_2$,
should provide an excellent test of these predictions.

We are grateful to O. K. Andersen for many stimulating discussions. This
work was supported by the Office of Naval Research.

\end{document}